\documentclass[aps,prl,reprint,groupedaddress]{revtex4-1}
\usepackage{graphicx}
\usepackage{amsmath}
\usepackage{amssymb}
\usepackage{amsthm}
\usepackage{mathrsfs}

\newtheorem{theorem}{Theorem}
\newtheorem{lemma}{Lemma}
\newcommand{\proj}[1]{\left|#1\right\rangle\!\!\left\langle #1\right|}
\newcommand{\ket}[1]{\left|#1\right\rangle}
\newcommand{\tracedist}[1]{\left\|#1\right\|_1}
\newcommand{\dens}[1]{\mathscr{D}(\mathcal{H}_{#1})}
\newcommand{\paren}[1]{\left(#1\right)}
\newcommand{\eq}[2]{\begin{equation} #2 \label{#1} \end{equation}}
\newcommand{\supp}{\text{supp}}
\newcommand{\lmin}[1]{\lambda_\text{min}\paren{#1}}
\def\cH{\mathcal{H}}
\def\cD{\mathcal{D}}
\newcommand{\be}{\begin{equation}} 
\newcommand{\ee}{\end{equation}} 
\newcommand{\eps}{{\varepsilon}}\newcommand{\vphi}{{\varphi}}
\newcommand{\tvphi}{{\widetilde{\varphi}}}
\newcommand{\bea}{\begin{eqnarray}}
\newcommand{\eea}{\end{eqnarray}}
\def\rank{\rm{rank}}
\def\tr{\rm{Tr}}
\def\reff#1{(\ref{#1})}

\begin{document}

\title{Bounding Polynomial Entanglement Measures for Mixed States}
\author{Samuel Rodriques}
\author{Nilanjana Datta$^{a}$}\email{n.datta@statslab.cam.ac.uk}
\author{Peter Love$^{b}$}\email{plove@haverford.edu}
\affiliation{$^{a}$~Statistical Laboratory, Centre for Mathematical Sciences, University of Cambridge, Wilberforce Road, Cambridge CB3 0WB, U.K.}
\affiliation{$^{b}$~Department of Physics, Haverford College, Haverford, Pennsylvania, 19041, USA}

\date{\today}

\begin{abstract}
We generalize the notion of the best separable approximation (BSA) and best W-class approximation (BWA) to arbitrary pure state entanglement measures, defining the best zero-$E$ approximation (BEA). We show that for any polynomial entanglement measure $E$, any mixed state $\rho$ admits at least one ``$S$-decomposition,'' i.e., a decomposition in terms of a mixed state on which $E$ is equal to zero, and a single additional pure state with (possibly) non-zero $E$. We show that the BEA is not in general the optimal $S$-decomposition from the point of view of bounding the entanglement of $\rho$, and describe an algorithm to construct the entanglement-minimizing $S$-decomposition for $\rho$ and place an upper bound on $E(\rho)$. When applied to the three-tangle, the cost of the algorithm is linear in the rank $d$ of the density matrix and has accuracy comparable to a steepest descent algorithm whose cost scales as $d^8 \log d$. We compare the upper bound to a lower bound algorithm given by Eltschka and Siewert for the three-tangle, and find that on random rank-two three-qubit density matrices, the difference between the upper and lower bounds is $0.14$ on average. We also find that the three-tangle of random full-rank three qubit density matrices is less than $0.023$ on average.
\end{abstract}


\maketitle

Non-classical correlations in quantum states such as entanglement distinguish quantum from classical information theory. The ability to calculate entanglement of mixed quantum states is relevant for the analysis of tomography data for systems of multiple qubits in several implementations~\cite{Volz06,RWZ05,HRB08}. Multipartite systems can contain multiple inequivalent types of entanglement that cannot be converted into one another by local operations and classical communication~\cite{DVC00}. 

One approach to characterizing pure-state entanglement in a system of qubits associates a polynomial function that is invariant under determinant $1$ local operations with each type of entanglement~\cite{EBOS12,MW02,Vidal00}. Examples of such polynomial invariants include the concurrence for two qubits~\cite{Wootters98} and the three-tangle, which quantifies the amount of entanglement  in a three-qubit system that cannot be accounted for by entanglement between pairs of the qubits~\cite{CKW00}.

A polynomial invariant $E$ is extended to mixed states by way of the convex roof, given for a rank-$d$ density matrix $\rho$ by:
\eq{ConvexRoof}{E(\rho) = \min_{\mathscr{E} \in \Upsilon_\rho}\sum_i p_i E({\psi_i}),}
where $\mathscr{E} = \{p_i, \ket{\psi_i}\}$ is a pure-state ensemble for $\rho$ and $\Upsilon_\rho$ is the set of all such ensembles. Caratheodory's theorem allows us to restrict the optimization to ensembles containing no more than $d^2$ elements~\cite{Uhlmann98}. 

An ensemble that minimizes eq.~\eqref{ConvexRoof} is said to be {\em{minimal}}. We consider the rank $d$ of the density matrix $d$, rather than the dimension of the Hilbert space on which it acts, because $d$ is the parameter that determines the computational difficulty of the convex roof minimization. A number of special cases of computation of the convex roof have been solved for cases of restricted rank~\cite{Hill97,EOSU08,LOSU06}.

Minimal ensembles have been found analytically for the concurrence of arbitrary two-qubit mixed states~\cite{Wootters98}, and for the three-tangle of rank-two mixtures of generalized GHZ and generalized W states~\cite{EOSU08,LOSU06}, as well as on rank-three mixtures of a GHZ state, a W state, and a state obtained by flipping all three bits of a W state~\cite{JHPS09}. When the minimal ensemble is not known analytically, which is the typical case, one may evaluate an upper bound on $E(\rho)$ using, for example, a steepest descent algorithm~\cite{ZKARBL12}. However, the cost of such an upper bound scales like $d^8 \log d$ making calculations infeasible for high rank.

An alternative approach to characterizing the entanglement of a mixed state was given by Lewenstein and Sanpera~\cite{LS98}. Given a two-qubit density matrix $\rho$, they they considered the set $S$ of pure states $\{\psi_i\}$, such that for all $\psi_i \in S$, there exists some $p_i \in (0,1)$ and a separable mixed state $\pi_i$ such that
\eq{SDecomp}{\rho = p_i \psi_i + (1-p_i) \pi_i.}
We refer to any decomposition of a state into  a pure state $\psi_i$ and a state $\pi_i$ such that $E(\pi_i)=0$, for given polynomial invariant $E$, as an $S$-decomposition. For the concurrence on two-qubit states, Lewenstein and Sanpera showed that $S$ is non-empty, and then considered the $S$-decompositions obtained by finding the element $\psi_e \in S$ that minimizes the corresponding probability $p_e$~\cite{LS98}. The corresponding separable state $\pi_e$ is the ``best separable approximation'' (BSA) of $\rho$. Their algorithm for finding the BSA of a mixed state $\rho$ determines whether $\rho$ is separable, and provides an upper bound on the entanglement of $\rho$, because $E(\rho) \leq p_e E(\psi_e)$ for all convex roof entanglement monotones $E$. 

More generally, it has been shown that every bipartite state $\rho$ has a unique convex decomposition of the form $\rho = \lambda \rho_s + (1-\lambda) \omega$, where $\rho_s$ is a separable state and the parameter $\lambda\in [0,1]$ is maximal~\cite{LS98, KL01}. The state $\rho_s$ is referred to as the BSA of $\rho$, and $\lambda$ as its {\em{separability}}~\cite{WK01}. Obviously, for any two-qubit entangled state, $\lambda < 1$. Moreover, it was established that for any two-qubit mixed state $\rho$, the separability is non-zero (and hence the BSA exists) and $\omega$ is a pure state. However, for states of more than two qubits, the separability may be zero and $\omega$ may be mixed. For a two-qubit state, the BSA places an upper bound on the concurrence, $C(\rho)$, of $\rho$, because clearly,
\eq{}{C(\rho) \leq (1-\lambda) C(\omega).}
All entanglement measures are zero on separable states. However, there are other interesting classes of states, such as the W-class, on which some, but not all, measures are zero. Acin et al. extended the above approach to the three-tangle of three-qubit states, defining the ``best W approximation'' (BWA)~\cite{ABLS01}.

In this paper, we generalize the BSA and BWA to arbitrary polynomial invariants, defining the best zero-$E$ approximation (BEA) of a mixed state $\rho$. We show that the BEA exists for all $\rho$ (i.e., $S$ is non-empty for all $\rho$) and that it is unique. For the BSA, BWA and BEA, the probability of the single entangled state in the pure-state ensemble for $\rho$ is minimized. This does not mean that the BSA, BEA or BWA gives the best upper bound on the entanglement for an ensemble of this form. One can obtain an improved upper bound on the entanglement of $\rho$ by finding the element $\psi_l$ of $S$ such that $p_l E(\psi_l)$ is minimized, where $p_l$ is the probability with which $\psi_l$ occurs in a convex decomposition of $\rho$. 

We describe an algorithm that finds $\psi_l$ for any mixed state $\rho$, and so places an upper bound on $E(\rho)$. Applied to the three-tangle, the cost of finding this upper bound scales linearly in $d$ and terminates after $10$ seconds on random three-qubit density matrices (using Intel Core $2$ CPUs at $2.66$ GHz), as opposed to the $d^8 \log d$ scaling and $10$ day runtime expected (see Table~\ref{TenIterationsComplexity}) for the steepest descent algorithm given in ~\cite{ZKARBL12}. We evaluate the accuracy of this algorithm by comparing this upper bound to the analytical value of the three-tangle for states on which it is known. By comparison of the upper bound and steepest descent methods on random states we demonstrate that the two algorithms exhibit comparable accuracies on states for which no analytical value is known. 

In all that follows, $\mathcal{H}$ denotes a Hilbert space of some number of qubits, $\dens{}$ denotes the set of density matrices (states) acting on $\mathcal{H}$, and $E: \dens{} \mapsto \mathbb{R}$ is assumed to be the convex roof extension of a polynomial function of pure states that is of homogeneous degree in the expansion coefficients of pure states written relative to the computational basis and that is invariant under determinant-$1$ local operations. $D(\rho,\pi) = \tracedist{\rho-\pi}$ is the trace distance, $\supp(\rho)$ is the support of 
$\rho$, and $R(\rho)$ is its range. For any pure state $|\psi\rangle \in \cH$, we denote the projector $|\psi\rangle \langle \psi|$ simply as $\psi$.

 We generalize the BSA and BWA, beginning with the following:
\begin{theorem}\label{thm1}
{{For any mixed state $\rho$ and polynomial invariant $E$ there exists a pure state ensemble containing at most one state with non-zero $E$.}}
\end{theorem}

The proof is given in the appendix. Theorem~\ref{thm1} leads naturally to an approximation of $\rho$ in terms of a mixed state for which $E$ is equal to zero. By analogy with the BSA and BWA, we define the BEA of $\rho$ as the state $\rho_e := \rho^*/\tr \rho^*$, where $\rho^* $ is a positive semi-definite operator with $E(\rho^*)=0$ such that $\rho - \rho^* \ge 0$ and $\tr \rho^* \le 1$ is maximal (Since $E(\rho^*)$ is a homogeneous polynomial in the expansion coefficients of the pure states in the minimal ensemble, it is well-defined even if $\rho^*$ has non-unit trace). Moreover, we refer to the parameter $\mu:=\tr \rho^* \in [0,1]$ as the {\em{zero-$E$ equivalency}} of $\rho$. Any state $\rho$ has a convex decomposition of the form
\be\label{wdecomp}
\rho = \mu \rho_e + (1-\mu) \omega,
\ee
where $\omega$ is a pure state with non-zero $E$. We refer to \reff{wdecomp} as the {\em{optimal zero-$E$ decomposition}} of $\rho$, and $\rho_e$ is the BEA. We now prove the following:

\begin{theorem}\label{thm2}
{All mixed states $\rho$ have non-zero zero-$E$ equivalency, and have a unique optimal zero-$E$ decomposition with $\omega$ being a pure state.}
\end{theorem}

Theorem~\ref{thm2} relies on Lemmas~\ref{lem1},~\ref{lem2} and \ref{lem3} whose proofs, together with the proofs of Theorems~\ref{thm1} and~\ref{thm2} are given in the Appendix.
\begin{lemma}\label{lem1}
Consider $\rho,\pi \in \dens{}$ and let $E:\dens{} \mapsto \mathbb{R}$ be a non-negative convex function bounded above by $E_\text{max}$. Suppose that there exists some $k>0$ such that
\eq{SigmaDefinitions}{\sigma_\rho = \rho + \frac{k}{D(\rho,\pi)}(\rho-\pi)}
is a state. Then,
\eq{MainClaim.1}{E(\rho)-E(\pi) \leq \frac{D(\rho,\pi)}{D(\sigma_\rho,\pi)}(E(\sigma_\rho)-E(\pi))}
\end{lemma}
The question of the existence of states of the form given by eq.~\eqref{SigmaDefinitions} is addressed by:
\begin{lemma}\label{lem2}
For all $\rho,\pi \in \cD(\cH)$ satisfying $\supp(\pi) \subseteq \supp(\rho)$, there exists a positive constant $k > 0$ such that the operator $\sigma_\rho$ defined as
\bea
\sigma_\rho &:=& \rho + \frac{k}{D(\rho,\pi)}(\rho-\pi)\label{SigmaDefinitionsBody}
\eea
is a state, and such that $\rank\,\sigma_\rho < \rank\,\rho$.
\end{lemma}
Eq.~\eqref{MainClaim.1}, combined with Lemma~\ref{lem2}, provides a non-uniform continuity bound on any non-negative convex function $E:\dens{} \mapsto \mathbb{R}$. The continuity bound is non-trivial between two density matrices $\rho$ and $\pi$ as long as $\rho$ and $\pi$ have equal supports.
 
We have now generalized the BSA and BWA to arbitrary homogeneous polynomial invariants. However, the BEA for $\rho$ does not in general provide the best estimate of $E(\rho)$ over the set of $S$-decompositions. The entanglement of the $S$-decomposition for $\rho$ (note that this is an upper bound on the entanglement of $\rho$ itself) with pure state $\psi$ occuring with probability $p$ is simply $p E(\psi)$.  Hence, we define $\psi_l$ to be the state in $S$ such that $p_l E(\psi_l)$ is minimal, and note
\eq{RhoBound}{E(\rho) \leq p_l E(\psi_l) \leq p_e E(\psi_e),}
where $\psi_e$ is the pure state associated with the BEA for $\rho$. Because the BEA is unique, the second inequality is only an equality if the BEA minimizes $p_e E(\psi_e)$ as well as $p_e$, i.e. if $p_e=p_l$ and $\psi_e=\psi_l$.

We now describe an algorithm that may be used to determine $\psi_l$, for any state $\rho$. We use the fact (from Lemma~\ref{lem3}) that every mixed state $\rho$ has in its range at least one pure state on which $E$ is equal to zero.
\begin{lemma}
\label{lem3}
For any mixed state $\rho$, there is a pure state 
$|\psi\rangle \in R(\rho)$ such that $E(\psi) = 0$.
\end{lemma}
Given a mixed state $\rho \in {\cal D}({\cal H})$ of rank $d$, we first use a steepest descent algorithm~\footnote{This algorithm, which performs steepest descent to minimize $E$ over $\mathcal{H}$ with cost independent of $d$, should not be confused with the steepest descent convex roof algorithm, which minimizes the three-tangle over $\Upsilon_\rho$ and which has cost scaling like $d^8 \log d$.} to identify pure states $\psi_i \in R(\rho)$ which have zero $E$~\footnote{These pure states could also be found using a root finding algorithm on the polynomial defined in the proof of Theorem~\ref{thm1}.}. For $d>2$ there is a continuous set of such states, and the steepest descent algorithm 
chooses one such state randomly.  We repeat this procedure several times to identify a number~\footnote{Caratheodory's theorem implies that pure-state ensembles of size $d^2$ are required in general in order to minimize the convex roof. For this reason, $\pi_i$ is constructed as a uniform mixture of $3/2 d-i$ distinct pure states selected from the range of $\rho_i$, each with zero $E$. Because $\sum_{i=0}^{d-2}(3/2 d-i) = d^2-1$ his ensures that $d^2$ pure states are selected in total, including the final pure state.} of such pure states
$\{\psi_i\}$.

We then construct the uniform mixture $\pi_1$ of the pure states identified by the steepest descent algorithm. Clearly, $\supp(\pi_1) \subseteq \supp(\rho)$ and $E(\pi_1) = 0$. Then, by Lemma~\ref{lem2}, there exists a $k >0$ such that the operator:
\eq{Rho1Def}{\rho_1 = \rho + \frac{k}{D(\rho,\pi_1)}(\rho-\pi_1)}
is a state, and such that $\rank\,\rho_1 < \rank\,\rho$. We then apply Lemma~\ref{lem1} with $\sigma_\rho \equiv \rho_1$ and $\pi \equiv \pi_1$ to obtain,
\eq{ConvexRoofBound.0}{E(\rho) - E(\pi_1) \leq \frac{D(\rho,\pi_1)}{D(\rho_1,\pi_1)} (E(\rho_1) - E(\pi_1)).}
Hence, because $E(\pi_1) = 0$,
\eq{ConvexRoofBound.1}{E(\rho) \leq \frac{D(\rho,\pi_1)}{D(\rho_1,\pi_1)}E(\rho_1).}
From eq.~\eqref{Rho1Def}, $\rho$ may be written as a convex combination of $\rho_1$ and $\pi_1$.

 If $\rho_1$ is a pure state, then $\rho$ may be written as a convex combination of the states $\psi$, (comprising $\pi_1$), which have zero $E$, and the state $\rho_1$, which may have non-zero $E$. We have thus identified a pure-state ensemble for $\rho$ containing at most one pure state with non-zero $E$, and the algorithm terminates since $E(\rho_1)$ can be calculated directly. 

If $\rho_1$ is not pure, the same procedure is applied to $\rho_1$. We find a density matrix $\pi_2$, such that $E(\pi_2)=0$ and $\supp(\pi_2) \subseteq \supp(\rho_1)$, and construct $\rho_2$ from it. The state $\rho$ can then be written as a convex combination of the pure states comprising $\pi_1$ and $\pi_2$, and the (possibly mixed) state $\rho_2$. Then,
\eq{ConvexRoofBound.2}{E(\rho_1) \leq \frac{D(\rho_1,\pi_2)}{D(\rho_2,\pi_2)}E(\rho_2).}
We can now combine eq.~\eqref{ConvexRoofBound.2} with eq.~\eqref{ConvexRoofBound.1} to obtain
\eq{ConvexRoofBound.3}{E(\rho) \leq \frac{D(\rho,\pi_1)}{D(\rho_1,\pi_1)}\frac{D(\rho_1,\pi_2)}{D(\rho_2,\pi_2)}E(\rho_2).}
The procedure is then repeated for $\rho_2$. The algorithm terminates when one arrives at a state $\rho_i$ which is pure, in which case $E(\rho_i)$ may be calculated directly. Because $\rank\,\rho_i<\rank\,\rho_{i-1}$ for all $i$, the algorithm is guaranteed to terminate, and we have
\eq{ConvexRoofBound.4}{E(\rho) \leq \frac{D(\rho,\pi_1)}{D(\rho_1,\pi_1)}\ldots\frac{D(\rho_{d-1},\pi_d)}{D(\rho_d,\pi_d)}E(\rho_d),}
where $\rho_d=\psi_d$ is pure. Note that at the $i$th step, we only need to find one pure state with $E=0$ in the range of $\rho_i$ for the algorithm to proceed.

The algorithm described above constructs an ensemble with the property that apart from $\psi_d$, every pure state in the ensemble has $E=0$. Thus, $\psi_d \in S$. To find $\psi_l$, we then use the fact that $S$ is a connected set, and that the number of local minima in $S$ is bounded above by the polynomial degree of $E$. The connectedness of $S$ follows from the connectedness of the set of mixed states with zero $E$. For many entanglement monotones of interest, the degree of $E$ is only $2$ or $4$~\cite{EBOS12}, so it follows that if we perform steepest descent to minimize the function $p E(\psi)$ over the set of pure states $\psi \in S$, where $p$ is the probability of $\psi$ in the $S$-decomposition, starting from the state $\rho_d=\psi_d$, we will converge to the global minimum $\psi_l$ with high probability. 

The real dimension of $S$ scales linearly in the rank of the density matrix, so the steepest descent is tractable. To perform this steepest descent in practice, for every state $\psi_i \in S$, we denote by $k_i \in (0,1)$ the smallest value such that for some mixed state $\pi_i$ with $E(\pi_i) = 0$, $\rho = k_i \psi_i + (1-k_i) \pi_i$. One can calculate $k_i$ using the algorithm of Lewenstein and Sanpera, suitably adapted to use pure states with zero $E$, rather than separable states~\cite{LS98}. If no such value of $k_i$ exists, we set $k_i = \infty$. One may then calculate $\psi_l$ by minimizing $k_i E(\psi_i)$ by steepest descent over $\mathcal{H}$.  

However, we have found that in many cases one may use eq.~\eqref{ConvexRoofBound.4} on its own to obtain a tight and computationally tractable numerical upper bound on $E(\rho)$. In this approach, one runs the algorithm up to eq.~\eqref{ConvexRoofBound.4} many times, getting a different result each time, and then takes the smallest of these results as an upper bound on $E(\rho)$. 

To investigate the efficacy of this method we performed calculations of the three-tangle for three-qubit mixed states. The three tangle is the simplest multipartite entanglement measure and in this case the BEA is the BWA. The combination of the upper bound obtained in this way from eq.~\eqref{ConvexRoofBound.4} with the lower bound of~\cite{ES12B} provides non-trivial upper and lower bounds on the three-tangle that one may compute rapidly on arbitrary states of three-qubits. We evaluate both bounds for mixtures of GHZ-class and W-class states~\cite{DVC00} where the three-tangle is known analytically~\cite{EOSU08,LOSU06}, and for random rank-$d$ density matrices. We compare the upper bound to analytical values of the three-tangle where available. We also compare to the steepest descent algorithm given in~\cite{ZKARBL12} and to the lower bound algorithm of Eltschka and Siewert~\cite{ES12B} for the square root of the three-tangle, whose square gives a lower bound on the three-tangle. Both algorithms are stochastic and so we repeat each one many times on a given state and use the best result. For the calculations described below our upper bound is the minimum value obtained by running our algorithm $200$ times on a given density matrix. The lower bound is the maximum value obtained after running the Eltschka and Siewert algorithm~\cite{ES12B} $1000$ times. 

\begin{figure}
\includegraphics[width=0.4\textwidth]{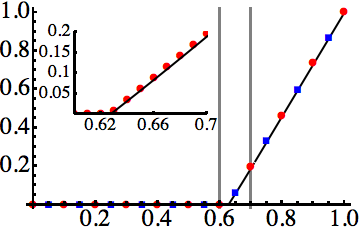}
\caption{\label{Analytics} Upper bounds on the three-tangle of the states in eq.~\eqref{GHZWMixture} for $11$ values of $p$ (red circles), compared to the analytical value (line). For ten values of $p$ (blue squares) the states in eq.~\eqref{GHZWMixture} were conjugated by a random element of $SU(2)^{\otimes 3}$ sampled from the Gaussian Unitary Ensemble. For $p=0$ through $p=0.6$, the algorithm calculated the three-tangle to be zero to numerical precision.  Inset: Upper bounds (red dots) on the three-tangle of the density matrices in eq.~\eqref{GHZWMixture} for $11$ values of $p$ between $0.6$ and $0.7$, compared to the analytical value (line). These are results from $400$ repetitions of the upper bound algorithm.}
\end{figure}

We evaluated the upper bound on mixtures of GHZ and W states,
\eq{GHZWMixture}{\pi(p) = p \proj{GHZ}+ (1-p) \proj{W},}
for which the analytical form is known~\cite{LOSU06}. Our algorithm was able to provide a tight upper bound for the three-tangle for this mixture (Fig.~\ref{Analytics}). In addition, whereas the steepest descent algorithm always yields a non-zero value for the three-tangle, the algorithm presented above can and does identify the three-tangle as exactly zero to within numerical precision if the final pure state $\rho_d$ has zero three-tangle. We also computed the upper bound for the case where the state is given by eq.~\eqref{GHZWMixture} convolved with a random element of $SU(2)^{\otimes 3}$ constructed from three elements of $SU(2)$ sampled independently from the Gaussian Unitary Ensemble. These states have identical entanglement to the states given by eq.~\eqref{GHZWMixture}. We performed $400$ repetitions of the algorithm and selected the smallest entanglement. Our findings are presented in Fig.~\ref{Analytics} and in Table~\ref{TenIterationsComplexity}.

We evaluated the upper bound on random mixtures of generalized GHZ and generalized W states for which the three tangle is also known analytically~\cite{LOSU06,EOSU08}. On $20000$ rank-two mixtures the average error of the upper-bound was $0.025$. On states that had zero three-tangle analytically, our upper bound yielded a value of exactly zero (to within numerical precision) $63\%$ of the time. The lower bound of~\cite{ES12B} was non-zero on $72\%$ of density matrices with non-zero three-tangle. 

The upper bound obtained from eq.~\eqref{ConvexRoofBound.4} was compared to the one obtained by the steepest descent algorithm described in~\cite{ZKARBL12} for random density matrices. For ranks $2$ through $5$, both algorithms calculated upper bounds on the three-tangle of $240$ different randomly generated density matrices.  For the algorithm presented above, density matrices of rank $6$, $7$ and $8$ were also tested.  The steepest descent algorithm yielded a lower (better) value on average but the difference decreased with increasing rank. The steepest descent was considerably slower than our algorithm for evaluating an upper bound, making calculations infeasible for ranks greater than $5$. The timings and average differences are shown in Table~\ref{TenIterationsComplexity}. 

\begin{table}
\begin{tabular}{c|ccccccc}
Rank & $2$ & $3$ & $4$ & $5$ & $6$ & $7$ & $8$\\ \hline
UB & $0.94$ & $1.89$ & $3.07$ & $4.2$ & $5.5$ & $7.0$ & $7.5$ \\
SD & $0.17$ & $1.39$ & $12.39$ & $125.2$ & & &\\
UB-SD&$0.0357$&$0.0239$&$0.0164$&$0.0116$&&&\\
\end{tabular}
\caption{Average runtimes (in seconds) for the algorithm presented in the text (UB) and the steepest descent algorithm (SD) on $240$ uniformly sampled three-qubit density matrices, for ranks two through eight. The steepest descent has only been calculated up to rank five due to the rapid growth in the runtime. The average difference in entanglements is shown in the third row (UB-SD).}
\label{TenIterationsComplexity}
\end{table}

\begin{figure}
\includegraphics[width=0.45\textwidth]{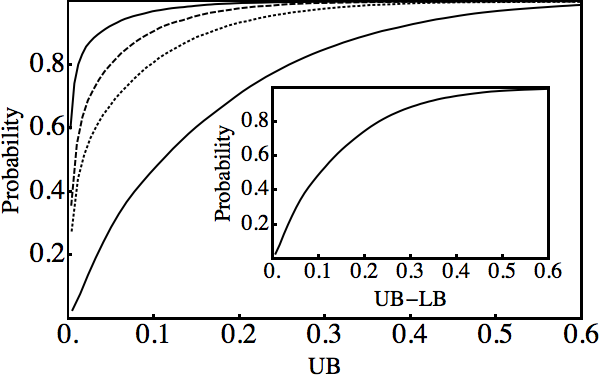}
\caption{\label{FigTwo} Cumulative distribution function of the upper bound on the three-tangle of $10000$ randomly sampled three-qubit density matrices of rank $2$ (lower solid line), $4$ (dotted line), $6$ (dashed line), $8$ (upper solid line). Inset: Cumulative distribution function of the difference between the upper and lower bounds on the three tangle of $30000$ uniformly sampled three-qubit rank $2$ density matrices.}
\end{figure}

We computed upper and lower bounds on random three-qubit states with ranks $2$ through $8$, for which the analytical form is not known. We sampled $30000$ states for rank $2$ and $10000$ states for ranks $3$ through $8$. We generated random rank $d$ states by sampling a probability distribution uniformly on the $(d-1)$-dimensional probability simplex and sampling pure states uniformly over the Hilbert space. . The upper bound tightly constrains the three-tangle in this ensemble of states, as shown in Figure~\ref{FigTwo}. The median values of the three-tangle for ranks $2$, $4$, $6$ and $8$ is $0.11$, $0.02$, $0.013$ and $0.003$ respectively. The lower bound was mostly zero on these states - only $2561$ of $30000$ states ($8.5\%$) were nonzero for rank $2$, $126$ of $10000$ for rank $3$, $12$ of $10000$ for rank $4$, $1$ of $10000$ for ranks $5$ and $6$ and none for ranks $7$ and $8$. Hence for random states of rank $>2$ the strategy of bounding the entanglement above and below is ineffective as we do not obtain a nontrivial lower bound from the method of~\cite{ES12B} in these cases.

For random rank-two states the mean upper and lower bounds over $30000$ states are $0.157$ and $0.016$, respectively, and the upper and lower bounds constrain the three-tangle to lie within a region of average width $0.14$. If we restrict to those states on which the lower bound is non-zero, so that we are considering states where we have a certificate that there is some entanglement, the mean upper and lower bounds are $0.356$ and $0.188$, respectively. Hence, states for which the lower bound is non-zero also have significantly larger values of the upper bound, and upper and lower bounds constrain the three-tangle to lie within a region of average width $0.167$ for these states. 

The algorithm will always terminate when applied to a polynomial entanglement monotone. On other convex roof entanglement monotones $E$, for which it may not be possible to construct ensembles containing at most one state on which $E$ is non-zero, the algorithm should choose $\pi_i$ to be the pure state in the support of $\rho_{i-1}$ on which $E$ is minimal. Then, eq.~\eqref{ConvexRoofBound.2} becomes (for the $(i-1)^\text{th}$ step)
\eq{ConvexRoofBound.5}{E(\rho_{i-1}) \leq \frac{D(\rho_{i-1},\pi_i)}{D(\rho_i,\pi_i)}\paren{E(\rho_i)-E(\pi_i)} + E(\pi_i).}
Since $\pi_i$ is pure, we may evaluate $E(\pi_i)$ analytically.

We generalized the best separable approximation (BSA) and best W-class approximation (BWA) to the best zero-E approximation (BEA) for any polynomial invariant. The BEA (like the BSA and BWA) is defined by minimizing the probability of the single entangled state in the ensemble that defines the BEA. We defined an algorithm that minimizes the entanglement for ensembles that contain a single entangled state. We have presented computations of upper and lower bounds for the three-tangle that are both practical methods for calculations on any three-qubit state. We validated these methods on mixtures of generalized GHZ and W states for which the exact value of the three-tangle is known. The upper and lower bounds are close on a large fraction of random rank-two states, including the fraction of those states for which the lower bound is non-zero and so for which we can certify that the entanglement is non-zero. Future work on the bounds may further close the gap between them and enable accurate estimation of the three-tangle, even if a closed form for the three-tangle remains out of reach.

\begin{acknowledgments}
SR would like to thank David Turban, Josh Sabloff, and PJL would like to thank Jens Siewert and David Meyer for useful discussions. ND is grateful to Anna Sanpera for helpful discussions.
\end{acknowledgments}

\section*{Appendix}

In this Appendix we give the proofs of the lemmas and theorems presented in 
the paper.
\begin{proof}[Proof of Lemma 1]
Let 
\be\label{omega}
\omega_p := p \pi + (1-p) \sigma_{\rho},
\ee
with $p \in (0,1)$. Since $\pi$ and $\sigma_\rho$ are states, so is $\omega_p$. Note that for the particular choice of $p$ given by
\be\label{pk}
p \equiv p_k = \frac{k}{D(\rho, \pi) + k},
\ee 
we have that $\omega_{p_k} = \rho$. Using the following expression for the trace distance between any two states $\rho_1$ and $\rho_2$:
$$ 
D(\rho_1, \rho_2) = \max_{0\le P \le I} \tr \left(P\left(\rho_1 - \rho_2\right)\right),
$$
it can be readily verified that the following identities hold:
\bea
D(\pi, \sigma_\rho) &=& D(\pi, \rho) + D(\rho, \sigma_\rho)\label{eqa}\\
{\hbox{and}} \quad D(\rho, \sigma_\rho) &=& k \label{eqb}.
\eea
Let $\eps:= 1 -p_k$. Then, \reff{pk}, \reff{eqa} and \reff{eqb} imply 
that 
\be
p_k = \frac{D(\rho, \sigma_\rho)}{D(\pi, \sigma_\rho)}
\quad {\hbox{and}} \quad \eps = \frac{D(\rho, \pi)}{D(\pi, \sigma_\rho)}\label{eps}.
\ee
Then,
\bea
E(\rho) - E(\pi) &=& E\left(\omega_{p_k}\right) -E\left(\omega_{p_k+\eps}\right)
\nonumber\\
&\le & {\eps}\left( E(\sigma_\rho) - E(\pi)\right)\nonumber\\
&=& \frac{D(\rho, \pi)}{D(\pi, \sigma_\rho)} \left( E(\sigma_\rho) - E(\pi)\right).
\label{eqbd}
\eea
The first equality holds since the choice of $p_k$ and $\eps$ ensures that
$\omega_{p_k} = \rho$ and $\omega_{p_k+\eps} = \omega_1= \pi$. The inequality in the second line holds since $E$ is a convex function and can be obtained as follows: Since $\eps = 1 - p_k$, we have $\omega_{p_k} = p_k \pi + \eps \sigma_\rho$. Then the convexity of $E$ implies that
$$ E(\omega_{p_k}) \le p_k E(\pi) + \eps E(\sigma_\rho),$$
and hence 
$$
E(\omega_{p_k}) - E(\pi) \le \eps\left(E(\sigma_\rho)) - E(\pi)\right).$$
The last equality in \reff{eqbd} follows from \reff{eps}.
\end{proof}

\begin{proof}[Proof of Lemma 2]
It is clear from \reff{SigmaDefinitions} that $\tr \sigma_\rho = 1$, since $\rho, \pi 
\in \cD(\cH)$. To establish that $\sigma_\rho$ is a state we only
need to show that $\sigma_\rho \ge 0$. 
In the following, for any $|\varphi\rangle \in \cH$ and any $\omega \in \cD(\cH)$ let 
$\omega^\varphi:= \langle \vphi | \omega |\vphi \rangle$. Any $|\vphi \rangle \in \cH$ can be written as
$$ |\vphi\rangle = \Pi_\rho |\vphi\rangle + (I - \Pi_\rho) |\vphi\rangle,$$
where $\Pi_\rho$ denotes the projection onto the support of $\rho$. Obviously, 
$\rho(I-\Pi_\rho) |\vphi\rangle=0$ and $\pi(I-\Pi_\rho) |\vphi\rangle=0$,
since ${\rm{supp}}\, (\pi) \subseteq 
{\rm{supp}}\,(\rho)$. These identities imply that $\sigma_\rho(I-\Pi_\rho) |\vphi\rangle=0$, and hence
$$\sigma_\rho^\vphi = \langle \vphi | \Pi_\rho \sigma \Pi_\rho |\vphi\rangle.$$
Let us define
\be
|\tvphi\rangle := \frac{\Pi_\rho |\vphi\rangle}{\sqrt{\langle \vphi |\Pi_\rho |\vphi\rangle}}.
\ee
Then to prove that $\sigma_\rho\ge 0$, it suffices to show that $\sigma_\rho^\tvphi \ge 0$.
From \eqref{SigmaDefinitions}, it equivalently suffices to prove that 
\be\label{step1}
\rho^\tvphi \ge \frac{k\left(\pi^\tvphi-\rho^\tvphi\right)}{D(\rho, \pi)}.
\ee
Note that $D(\rho,\pi) = D(\pi,\rho)$ by symmetry, and that
\bea
D(\pi, \rho) &=& \max_{0 \le P \le I}\tr \left(P(\pi-\rho)\right),\nonumber\\
&\ge & \tr \left(|\tvphi\rangle \langle \tvphi|(\pi-\rho) \right)\nonumber\\
&=& \pi^\tvphi-\rho^\tvphi.
\eea              
Hence to prove \reff{step1}, it suffices
to prove that there exists a positive constant $k$ such that 
\be\rho^\tvphi \ge k
\ee
Let the eigenvalue decomposition of $\rho$ be given by
\be\rho = \sum_{i=1}^d \lambda_i |e_i\rangle \langle e_i|,\ee
and let us choose $k=\lambda_{\min}(\rho)$, where
$\lambda_{\min}(\rho) := \min_{1\le i \le d} \{ \lambda_i \,:\, \lambda_i >0\}$.
Obviously $|\tvphi\rangle \in {\rm{supp }}\,\rho$ and hence 
$|\tvphi\rangle = \sum_{i: \lambda_i >0} \alpha_i |e_i\rangle,$
with $\alpha_i= \langle e_i |\tvphi \rangle$, and $\sum_{i: \lambda_i >0} |\alpha_i|^2=1.$ Hence 
\bea\label{Lmin}
\rho^\tvphi &=& \langle \tvphi | (\sum_{i : \lambda_i >0} \lambda_i |e_i \rangle \langle e_i|)|\tvphi\rangle\nonumber\\
&\ge & \lambda_{\min}(\rho) \sum_{i : \lambda_i >0} |\langle \tvphi| e_i \rangle |^2\nonumber\\
&=&  \lambda_{\min}(\rho).
\eea
It follows that the operator $\sigma_\rho$ defined in eq.~\eqref{SigmaDefinitions} with $k=\lmin{\rho}$ is a state. However, if $\rho \neq \pi$, then $\rho - \pi$ has at least one negative eigenvalue. It follows that, for $k>0$ large enough, the operator defined in eq.~\eqref{SigmaDefinitions} is not positive semi-definite, and hence is not a state. By continuity, then, there exists a $p>0$ such the operator $\sigma_\rho$ defined in eq.~\eqref{SigmaDefinitions} with $k=p$ is a state, but such that the operator $\sigma_\rho$ as defined in eq.~\eqref{SigmaDefinitions} is not a state when $k=p+\epsilon$, for any positive $\epsilon$. Define $\bar{\sigma}_\rho$ to be the operator $\sigma_\rho$ defined in eq.~\eqref{SigmaDefinitions} with $k=p$.

If the support of $\pi$ were contained within the support of $\bar{\sigma}_\rho$, then there would exist some $q>0$ such that
\eq{Positivity}{\bar{\sigma}_\rho + \frac{q}{D(\bar{\sigma}_\rho,\pi)} (\bar{\sigma}_\rho - \pi) \geq 0.}
If such a $q$ existed, however, then one could use the fact that, from the definitions of $\sigma_\rho$ and the trace distance,
\[p=D(\rho,\sigma_\rho)\]
and
\[D(\pi,\rho) + D(\rho,\sigma_\rho) = D(\pi, \sigma_\rho),\]
in order to show that
\[\rho + \frac{p+q}{D(\rho,\pi)} (\rho - \pi) \geq 0,\]
which is a contradiction by the definition of $p$. Thus, the support of $\pi$ is not contained within the support of $\bar{\sigma}_\rho$, so the supports of $\bar{\sigma}_\rho$ and $\rho$ are not equal. However, $\rho$ can be written as a convex combination of $\bar{\sigma}_\rho$ and $\pi$, so the support of $\bar{\sigma}_\rho$ must be contained within the support of $\rho$. It follows that the support of $\bar{\sigma}_\rho$ is strictly smaller than the support of $\rho$, so the rank of $\bar{\sigma}_\rho$ must be smaller than the rank of $\rho$.
\end{proof}

\begin{proof}[Proof of Lemma 3]
Since the range of any mixed state $\rho$ is the set of superpositions of the eigenvectors of $\rho$, it suffices to show that for any two pure states $\ket{\psi_1}$ and $\ket{\psi_2}$, there exists some $\theta$, $\phi$, such that the state
\eq{}{\ket{\theta,\phi} = \cos(\theta) \ket{\psi_1} + e^{i \phi} \sin(\theta) \ket{\psi_2}}
has zero $E$.

Let the coefficients of $\ket{\psi_1}$ in the computational basis be written $b_1,b_2,\ldots,b_8$, and let the coefficients of $\ket{\psi_2}$ in the computational basis be written $c_1,c_2,\ldots,c_8$. Then for all $k$, the coefficients of $\ket{\theta,\phi}$, written in the computational basis, are given by
\eq{f.0}{a_k = \cos(\theta) b_k + e^{i \phi} \sin(\theta) c_k.}
Since $E$ is a polynomial of homogeneous degree $D$ at most 4 in the coefficients of $\ket{\theta,\phi}$ written in the computational basis, we may write it as
\eq{f.1}{E(\ket{\theta,\phi})=\sum_{i}^T C_i a_1^{s_{1,i}} \ldots a_N^{s_{N,i}},}
where $T$ is the number of terms in the expression for $E$, $N$ is the dimension of the Hilbert space, $C_i$ is the coefficient on the $i$th term, and $\sum_{k=1}^{N} s_{k,i} = D$ for all $i$. Substituting the expression for $a_k$, we have
\eq{f.2}{E(\ket{\theta,\phi})=\sum_{i}^T C_i \prod_{k=1}^N \paren{ \cos(\theta) b_k + e^{i \phi} \sin(\theta) c_k}^{s_{k,i}}.}
Factoring out $\cos(\theta)^D$ from every term in the sum, we have,
\eq{f.3}{E(\ket{\theta,\phi})=\cos(\theta)^D\sum_{i}^T C_i \prod_{k=1}^N \paren{b_k + e^{i \phi} \tan(\theta) c_k}^{s_{k,i}}.}
We now perform a change of variables, defining $z(\theta,\phi) = e^{i \phi} \tan(\theta)$. We have
\eq{f.4}{E(\ket{\theta,\phi})=\cos(\theta)^D\sum_{i}^T C_i \prod_{k=1}^N \paren{b_k + z(\theta,\phi) c_k}^{s_{k,i}}.}
We now note that the range of $\theta$ may be restricted to the interval $[0,\pi/2]$, while the range of $\phi$ is $[0,2\pi]$. If we assume that $E(\ket{\psi_2}) \neq 0$, then for the purpose of finding the roots of $E$, the range of $\theta$ may be restricted further to $[0,\pi/2)$. On this range, $\cos(\theta)^D$ is non-zero, so the roots of $E(\ket{\theta,\phi}) = 0$ are equivalent to the roots of the polynomial
\eq{f.5}{\bar{E}=\sum_{i}^T C_i \prod_{k=1}^N \paren{b_k + z(\theta,\phi) c_k}^{s_{k,i}}.}
The fundamental theorem of algebra guarantees that $\bar{E}$ will have $D$ complex roots, including multiplicities. These roots lie within the range of $z(\theta,\phi)$, which is the entire complex plane. Thus, there exists at least one unique pure state $\ket{\theta,\phi}$ such that $E(\ket{\theta,\phi})=0$, completing the proof.
\end{proof}

\begin{proof}[Proof of Theorem 2]
 
Every pure state $|\psi\rangle \in R(\omega)$ (where $\omega$ is the state appearing in the optimal zero-$E$ decomposition~\eqref{wdecomp}) must have positive $E$. This is because, if there is a pure state $|\varphi\rangle \in R(\omega)$ with $E(\varphi) = 0$, then we could subtract $\gamma \varphi$
from $\omega$ (for some $0<\gamma <1$) and add $(1-\mu) \gamma \varphi$ to $\mu \rho_L$ to obtain a decomposition of the form:
$$\rho = \tilde{\mu} {\tilde{\rho}}_L + (1 -  \tilde{\mu}) \tilde{\omega},
$$
such that $E( {\tilde{\rho}}_L)=0$ and $\tilde{\mu} = \mu + (1-\mu)\lambda$. However, this increases the zero-$E$ equivalency $\mu$, and hence leads to a contradiction, since $\mu$ is maximal by definition. Hence $E(\psi)>0$ $\forall \psi \in R(\omega)$. By Lemma~3, it then follows that $\omega$ must be pure. This also implies that if $\rho$ is a mixed state, then $\mu>0$.

To prove that the optimal zero-$E$ decomposition is unique, we proceed as in~\cite{KL01} and assume that there
exists at least two optimal zero-$E$ decompositions $\rho= \lambda \rho_L + (1-\lambda) |\psi\rangle \langle \psi|$ and  $\rho= \lambda \rho_L^\prime + (1-\lambda) |\psi'\rangle \langle \psi'|$, with the same maximal $\lambda$. Any convex
combination of these two decompositions is also an optimal zero-$E$ decomposition, i.e., $\forall$ $\eps \in [0,1]$,
\bea
\rho &=& \eps ( \lambda \rho_L + (1-\lambda) \psi) +
(1- \eps)  (\lambda \rho_L^\prime + (1-\lambda) \psi')\nonumber\\
&=& \lambda {\tilde{\rho}}_L + (1-\lambda) \tilde{\omega},
\eea
where $\tilde{\omega} := \eps \psi + (1-\eps)\psi'$ and $ {\tilde{\rho}}_L := \eps  \rho_L + (1-\eps)\rho_L^\prime$,
with $E( {\tilde{\rho}}_L)=0$ (since the convex roof extension $E$ is convex and 
$E( {{\rho}}_L)=0=E( {{\rho}}_L^\prime)=0$).  Since $\tilde{\omega}$ is a mixed state, by Lemma~3 there must exist a pure state $|\varphi>$ in its range
such that (as above) we could subtract $c \varphi$ from $\tilde{\omega}$ (for some $c\in (0,1)$) and add it to $\lambda {\tilde{\rho}}_L$ to obtain another optimal zero-$E$ decomposition. However, this would increase the zero-$E$ equivalency and thus result in a contradiction.
\end{proof}

\begin{proof}[Proof of Theorem 1]
We prove Theorem 1 by construction, using Lemmas 1,2 and 3. Since $\rho$ is a mixed state, by Lemma 3 there exists a pure state $|\psi\rangle \in R(\rho)$ such that $E(\psi) = 0$. By Lemma 2 we infer that there exists a positive constant $k$ and a state $\rho_1$ such that 
\be\label{eq1}
\rho = \lambda \psi + (1-\lambda) \rho_1,
\ee
with $\rank \,\rho_1 < \rank \,\rho$. Here $\lambda \equiv \lambda(k,D(\rho, \psi))= k/(D(\rho, \psi) + k)$.

If $\rho_1$ is a pure state, then the claim of Theorem 1 follows, since we have constructed an ensemble of two pure states $\psi$ and $\rho_1$ for $\rho$, with $E(\psi)=0$ and $E(\rho_1) \ge 0$. If $\rho_1$ is a mixed state, then we know by Lemma 2 that there exists a pure state $|\psi_1\rangle \in 
R(\rho_1)$ such  that $E(\psi_1)=0$. Then we repeat the above steps (for $\rho_1$) to arrive at a state $\rho_2$ such that 
$$\rho = \lambda_1 \psi_1 + (1-\lambda_1) \rho_2,$$
and $\lambda_1 \in (0,1)$. If $\rho_2$ is pure, the proof is completed since 
$\rho$ can be expressed in terms of an ensemble of three pure states $\psi, \psi_1$ and $\rho_2$, with only $\rho_2$ having possibly non-zero $E$. If $\rho_2$ is mixed, we iterate again. We stop after the $i^{th}$ iteration if $\rho_i$ is pure. Since, by Lemma 2, the rank of the state $\rho_i$ obtained after the ${i^{th}}$ iteration is strictly smaller than the rank of $\rho_{i+1}$, we definitely arrive at a pure state and hence the iteration stops, yielding an ensemble of pure states for $\rho$ with at most one (namely the one obtained in the last step of the iteration) having non-zero $E$.
\end{proof}


\begin{thebibliography}{23}%
\makeatletter
\providecommand \@ifxundefined [1]{%
 \@ifx{#1\undefined}
}%
\providecommand \@ifnum [1]{%
 \ifnum #1\expandafter \@firstoftwo
 \else \expandafter \@secondoftwo
 \fi
}%
\providecommand \@ifx [1]{%
 \ifx #1\expandafter \@firstoftwo
 \else \expandafter \@secondoftwo
 \fi
}%
\providecommand \natexlab [1]{#1}%
\providecommand \enquote  [1]{``#1''}%
\providecommand \bibnamefont  [1]{#1}%
\providecommand \bibfnamefont [1]{#1}%
\providecommand \citenamefont [1]{#1}%
\providecommand \href@noop [0]{\@secondoftwo}%
\providecommand \href [0]{\begingroup \@sanitize@url \@href}%
\providecommand \@href[1]{\@@startlink{#1}\@@href}%
\providecommand \@@href[1]{\endgroup#1\@@endlink}%
\providecommand \@sanitize@url [0]{\catcode `\\12\catcode `\$12\catcode
  `\&12\catcode `\#12\catcode `\^12\catcode `\_12\catcode `\%12\relax}%
\providecommand \@@startlink[1]{}%
\providecommand \@@endlink[0]{}%
\providecommand \url  [0]{\begingroup\@sanitize@url \@url }%
\providecommand \@url [1]{\endgroup\@href {#1}{\urlprefix }}%
\providecommand \urlprefix  [0]{URL }%
\providecommand \Eprint [0]{\href }%
\providecommand \doibase [0]{http://dx.doi.org/}%
\providecommand \selectlanguage [0]{\@gobble}%
\providecommand \bibinfo  [0]{\@secondoftwo}%
\providecommand \bibfield  [0]{\@secondoftwo}%
\providecommand \translation [1]{[#1]}%
\providecommand \BibitemOpen [0]{}%
\providecommand \bibitemStop [0]{}%
\providecommand \bibitemNoStop [0]{.\EOS\space}%
\providecommand \EOS [0]{\spacefactor3000\relax}%
\providecommand \BibitemShut  [1]{\csname bibitem#1\endcsname}%
\let\auto@bib@innerbib\@empty
%</preamble>
\bibitem [{\citenamefont {Volz}\ \emph {et~al.}(2006)\citenamefont {Volz},
  \citenamefont {Weber}, \citenamefont {Schlenk}, \citenamefont {Rosenfeld},
  \citenamefont {Vrana}, \citenamefont {Saucke}, \citenamefont {Kurtsiefer},\
  and\ \citenamefont {Weinfurter}}]{Volz06}%
  \BibitemOpen
  \bibfield  {author} {\bibinfo {author} {\bibfnamefont {J.}~\bibnamefont
  {Volz}}, \bibinfo {author} {\bibfnamefont {M.}~\bibnamefont {Weber}},
  \bibinfo {author} {\bibfnamefont {D.}~\bibnamefont {Schlenk}}, \bibinfo
  {author} {\bibfnamefont {W.}~\bibnamefont {Rosenfeld}}, \bibinfo {author}
  {\bibfnamefont {J.}~\bibnamefont {Vrana}}, \bibinfo {author} {\bibfnamefont
  {K.}~\bibnamefont {Saucke}}, \bibinfo {author} {\bibfnamefont
  {C.}~\bibnamefont {Kurtsiefer}}, \ and\ \bibinfo {author} {\bibfnamefont
  {H.}~\bibnamefont {Weinfurter}},\ }\href@noop {} {\bibfield  {journal}
  {\bibinfo  {journal} {Phys. Rev. Lett.}\ }\textbf {\bibinfo {volume} {96}},\
  \bibinfo {pages} {030404} (\bibinfo {year} {2006})}\BibitemShut {NoStop}%
\bibitem [{\citenamefont {Resch}\ \emph {et~al.}(2005)\citenamefont {Resch},
  \citenamefont {Walther},\ and\ \citenamefont {Zeilinger}}]{RWZ05}%
  \BibitemOpen
  \bibfield  {author} {\bibinfo {author} {\bibfnamefont {K.}~\bibnamefont
  {Resch}}, \bibinfo {author} {\bibfnamefont {P.}~\bibnamefont {Walther}}, \
  and\ \bibinfo {author} {\bibfnamefont {A.}~\bibnamefont {Zeilinger}},\
  }\href@noop {} {\bibfield  {journal} {\bibinfo  {journal} {Phys. Rev. Lett.}\
  }\textbf {\bibinfo {volume} {94}},\ \bibinfo {pages} {070402} (\bibinfo
  {year} {2005})}\BibitemShut {NoStop}%
\bibitem [{\citenamefont {H{\"a}ffner}\ \emph {et~al.}(2008)\citenamefont
  {H{\"a}ffner}, \citenamefont {Roos},\ and\ \citenamefont {Blatt}}]{HRB08}%
  \BibitemOpen
  \bibfield  {author} {\bibinfo {author} {\bibfnamefont {H.}~\bibnamefont
  {H{\"a}ffner}}, \bibinfo {author} {\bibfnamefont {C.~F.}\ \bibnamefont
  {Roos}}, \ and\ \bibinfo {author} {\bibfnamefont {R.}~\bibnamefont {Blatt}},\
  }\href@noop {} {\bibfield  {journal} {\bibinfo  {journal} {Phys. Rep.}\
  }\textbf {\bibinfo {volume} {469}},\ \bibinfo {pages} {155} (\bibinfo {year}
  {2008})}\BibitemShut {NoStop}%
\bibitem [{\citenamefont {D{\"u}r}\ \emph {et~al.}(2000)\citenamefont
  {D{\"u}r}, \citenamefont {Vidal},\ and\ \citenamefont {Cirac}}]{DVC00}%
  \BibitemOpen
  \bibfield  {author} {\bibinfo {author} {\bibfnamefont {W.}~\bibnamefont
  {D{\"u}r}}, \bibinfo {author} {\bibfnamefont {G.}~\bibnamefont {Vidal}}, \
  and\ \bibinfo {author} {\bibfnamefont {J.~I.}\ \bibnamefont {Cirac}},\
  }\href@noop {} {\bibfield  {journal} {\bibinfo  {journal} {Phys. Rev. A}\
  }\textbf {\bibinfo {volume} {62}},\ \bibinfo {pages} {062314} (\bibinfo
  {year} {2000})}\BibitemShut {NoStop}%
\bibitem [{\citenamefont {Eltschka}\ \emph {et~al.}(2012)\citenamefont
  {Eltschka}, \citenamefont {Bastin}, \citenamefont {Osterloh},\ and\
  \citenamefont {Siewert}}]{EBOS12}%
  \BibitemOpen
  \bibfield  {author} {\bibinfo {author} {\bibfnamefont {C.}~\bibnamefont
  {Eltschka}}, \bibinfo {author} {\bibfnamefont {T.}~\bibnamefont {Bastin}},
  \bibinfo {author} {\bibfnamefont {A.}~\bibnamefont {Osterloh}}, \ and\
  \bibinfo {author} {\bibfnamefont {J.}~\bibnamefont {Siewert}},\ }\href@noop
  {} {\bibfield  {journal} {\bibinfo  {journal} {Physical Review A}\ }\textbf
  {\bibinfo {volume} {85}},\ \bibinfo {pages} {022301} (\bibinfo {year}
  {2012})}\BibitemShut {NoStop}%
\bibitem [{\citenamefont {Meyer}\ and\ \citenamefont {Wallach}(2002)}]{MW02}%
  \BibitemOpen
  \bibfield  {author} {\bibinfo {author} {\bibfnamefont {D.~A.}\ \bibnamefont
  {Meyer}}\ and\ \bibinfo {author} {\bibfnamefont {N.~R.}\ \bibnamefont
  {Wallach}},\ }\href@noop {} {\bibfield  {journal} {\bibinfo  {journal} {J.
  Math. Phys.}\ }\textbf {\bibinfo {volume} {43}},\ \bibinfo {pages} {4273}
  (\bibinfo {year} {2002})}\BibitemShut {NoStop}%
\bibitem [{\citenamefont {Vidal}(2000)}]{Vidal00}%
  \BibitemOpen
  \bibfield  {author} {\bibinfo {author} {\bibfnamefont {G.}~\bibnamefont
  {Vidal}},\ }\href@noop {} {\bibfield  {journal} {\bibinfo  {journal} {J. Mod.
  Opt.}\ }\textbf {\bibinfo {volume} {47}},\ \bibinfo {pages} {355} (\bibinfo
  {year} {2000})}\BibitemShut {NoStop}%
\bibitem [{\citenamefont {Wootters}(1998)}]{Wootters98}%
  \BibitemOpen
  \bibfield  {author} {\bibinfo {author} {\bibfnamefont {W.}~\bibnamefont
  {Wootters}},\ }\href@noop {} {\bibfield  {journal} {\bibinfo  {journal}
  {Phys. Rev. Lett.}\ }\textbf {\bibinfo {volume} {80}},\ \bibinfo {pages}
  {2245} (\bibinfo {year} {1998})}\BibitemShut {NoStop}%
\bibitem [{\citenamefont {Coffman}\ \emph {et~al.}(2000)\citenamefont
  {Coffman}, \citenamefont {Kundu},\ and\ \citenamefont {Wootters}}]{CKW00}%
  \BibitemOpen
  \bibfield  {author} {\bibinfo {author} {\bibfnamefont {V.}~\bibnamefont
  {Coffman}}, \bibinfo {author} {\bibfnamefont {J.}~\bibnamefont {Kundu}}, \
  and\ \bibinfo {author} {\bibfnamefont {W.~K.}\ \bibnamefont {Wootters}},\
  }\href {\doibase 10.1103/PhysRevA.61.052306} {\bibfield  {journal} {\bibinfo
  {journal} {Phys. Rev. A}\ }\textbf {\bibinfo {volume} {61}},\ \bibinfo
  {pages} {052306} (\bibinfo {year} {2000})}\BibitemShut {NoStop}%
\bibitem [{\citenamefont {Uhlmann}(1998)}]{Uhlmann98}%
  \BibitemOpen
  \bibfield  {author} {\bibinfo {author} {\bibfnamefont {A.}~\bibnamefont
  {Uhlmann}},\ }\href {http://dx.doi.org/10.1023/A:1009664331611} {\bibfield
  {journal} {\bibinfo  {journal} {Open Systems \& Information Dynamics}\
  }\textbf {\bibinfo {volume} {5}},\ \bibinfo {pages} {209} (\bibinfo {year}
  {1998})},\ \bibinfo {note} {10.1023/A:1009664331611}\BibitemShut {NoStop}%
\bibitem [{\citenamefont {Hill}\ and\ \citenamefont {Wootters}(1997)}]{Hill97}%
  \BibitemOpen
  \bibfield  {author} {\bibinfo {author} {\bibfnamefont {S.}~\bibnamefont
  {Hill}}\ and\ \bibinfo {author} {\bibfnamefont {W.~K.}\ \bibnamefont
  {Wootters}},\ }\href@noop {} {\bibfield  {journal} {\bibinfo  {journal}
  {Physical Review Letters}\ }\textbf {\bibinfo {volume} {78}},\ \bibinfo
  {pages} {5022} (\bibinfo {year} {1997})}\BibitemShut {NoStop}%
\bibitem [{\citenamefont {Eltschka}\ \emph {et~al.}(2008)\citenamefont
  {Eltschka}, \citenamefont {Osterloh}, \citenamefont {Siewert},\ and\
  \citenamefont {Uhlmann}}]{EOSU08}%
  \BibitemOpen
  \bibfield  {author} {\bibinfo {author} {\bibfnamefont {C.}~\bibnamefont
  {Eltschka}}, \bibinfo {author} {\bibfnamefont {A.}~\bibnamefont {Osterloh}},
  \bibinfo {author} {\bibfnamefont {J.}~\bibnamefont {Siewert}}, \ and\
  \bibinfo {author} {\bibfnamefont {A.}~\bibnamefont {Uhlmann}},\ }\href@noop
  {} {\bibfield  {journal} {\bibinfo  {journal} {New J. Phys.}\ }\textbf
  {\bibinfo {volume} {10}},\ \bibinfo {pages} {043014} (\bibinfo {year}
  {2008})}\BibitemShut {NoStop}%
\bibitem [{\citenamefont {Lohmayer}\ \emph {et~al.}(2006)\citenamefont
  {Lohmayer}, \citenamefont {Osterloh}, \citenamefont {Siewert},\ and\
  \citenamefont {Uhlmann}}]{LOSU06}%
  \BibitemOpen
  \bibfield  {author} {\bibinfo {author} {\bibfnamefont {R.}~\bibnamefont
  {Lohmayer}}, \bibinfo {author} {\bibfnamefont {A.}~\bibnamefont {Osterloh}},
  \bibinfo {author} {\bibfnamefont {J.}~\bibnamefont {Siewert}}, \ and\
  \bibinfo {author} {\bibfnamefont {A.}~\bibnamefont {Uhlmann}},\ }\href@noop
  {} {\bibfield  {journal} {\bibinfo  {journal} {Phys. Rev. Lett.}\ }\textbf
  {\bibinfo {volume} {97}},\ \bibinfo {pages} {260502} (\bibinfo {year}
  {2006})}\BibitemShut {NoStop}%
\bibitem [{\citenamefont {Jung}\ \emph {et~al.}(2009)\citenamefont {Jung},
  \citenamefont {Hwang}, \citenamefont {Park},\ and\ \citenamefont
  {Son}}]{JHPS09}%
  \BibitemOpen
  \bibfield  {author} {\bibinfo {author} {\bibfnamefont {E.}~\bibnamefont
  {Jung}}, \bibinfo {author} {\bibfnamefont {M.-R.}\ \bibnamefont {Hwang}},
  \bibinfo {author} {\bibfnamefont {D.}~\bibnamefont {Park}}, \ and\ \bibinfo
  {author} {\bibfnamefont {J.-W.}\ \bibnamefont {Son}},\ }\href {\doibase
  10.1103/PhysRevA.79.024306} {\bibfield  {journal} {\bibinfo  {journal} {Phys.
  Rev. A}\ }\textbf {\bibinfo {volume} {79}},\ \bibinfo {pages} {024306}
  (\bibinfo {year} {2009})}\BibitemShut {NoStop}%
\bibitem [{\citenamefont {Zhu}\ \emph {et~al.}(2012)\citenamefont {Zhu},
  \citenamefont {Kais}, \citenamefont {Aspuru-Guzik}, \citenamefont
  {Rodriques}, \citenamefont {Brock},\ and\ \citenamefont {Love}}]{ZKARBL12}%
  \BibitemOpen
  \bibfield  {author} {\bibinfo {author} {\bibfnamefont {J.}~\bibnamefont
  {Zhu}}, \bibinfo {author} {\bibfnamefont {S.}~\bibnamefont {Kais}}, \bibinfo
  {author} {\bibfnamefont {A.}~\bibnamefont {Aspuru-Guzik}}, \bibinfo {author}
  {\bibfnamefont {S.}~\bibnamefont {Rodriques}}, \bibinfo {author}
  {\bibfnamefont {B.}~\bibnamefont {Brock}}, \ and\ \bibinfo {author}
  {\bibfnamefont {P.~J.}\ \bibnamefont {Love}},\ }\href {\doibase
  10.1063/1.4742333} {\bibfield  {journal} {\bibinfo  {journal} {J. Chem.
  Phys.}\ }\textbf {\bibinfo {volume} {137}},\ \bibinfo {eid} {074112}
  (\bibinfo {year} {2012})}\BibitemShut {NoStop}%
\bibitem [{\citenamefont {Lewenstein}\ and\ \citenamefont
  {Sanpera}(1998)}]{LS98}%
  \BibitemOpen
  \bibfield  {author} {\bibinfo {author} {\bibfnamefont {M.}~\bibnamefont
  {Lewenstein}}\ and\ \bibinfo {author} {\bibfnamefont {A.}~\bibnamefont
  {Sanpera}},\ }\href@noop {} {\bibfield  {journal} {\bibinfo  {journal} {Phys.
  Rev. Lett.}\ }\textbf {\bibinfo {volume} {80}},\ \bibinfo {pages} {2261}
  (\bibinfo {year} {1998})}\BibitemShut {NoStop}%
\bibitem [{\citenamefont {Karnas}\ and\ \citenamefont
  {Lewenstein}(2001)}]{KL01}%
  \BibitemOpen
  \bibfield  {author} {\bibinfo {author} {\bibfnamefont {S.}~\bibnamefont
  {Karnas}}\ and\ \bibinfo {author} {\bibfnamefont {M.}~\bibnamefont
  {Lewenstein}},\ }\href@noop {} {\bibfield  {journal} {\bibinfo  {journal} {J.
  Phys. A}\ }\textbf {\bibinfo {volume} {34}},\ \bibinfo {pages} {6919}
  (\bibinfo {year} {2001})}\BibitemShut {NoStop}%
\bibitem [{\citenamefont {{Wellens}}\ and\ \citenamefont
  {{Ku{\'s}}}(2001)}]{WK01}%
  \BibitemOpen
  \bibfield  {author} {\bibinfo {author} {\bibfnamefont {T.}~\bibnamefont
  {{Wellens}}}\ and\ \bibinfo {author} {\bibfnamefont {M.}~\bibnamefont
  {{Ku{\'s}}}},\ }\href {\doibase 10.1103/PhysRevA.64.052302} {\bibfield
  {journal} {\bibinfo  {journal} {Phys. Rev. A}\ }\textbf {\bibinfo {volume}
  {64}},\ \bibinfo {pages} {052302} (\bibinfo {year} {2001})}\BibitemShut
  {NoStop}%
\bibitem [{\citenamefont {Acin}\ \emph {et~al.}(2001)\citenamefont {Acin},
  \citenamefont {Bru{\ss}}, \citenamefont {Lewenstein},\ and\ \citenamefont
  {Sanpera}}]{ABLS01}%
  \BibitemOpen
  \bibfield  {author} {\bibinfo {author} {\bibfnamefont {A.}~\bibnamefont
  {Acin}}, \bibinfo {author} {\bibfnamefont {D.}~\bibnamefont {Bru{\ss}}},
  \bibinfo {author} {\bibfnamefont {M.}~\bibnamefont {Lewenstein}}, \ and\
  \bibinfo {author} {\bibfnamefont {A.}~\bibnamefont {Sanpera}},\ }\href@noop
  {} {\bibfield  {journal} {\bibinfo  {journal} {Physical Review Letters}\
  }\textbf {\bibinfo {volume} {87}},\ \bibinfo {pages} {040401} (\bibinfo
  {year} {2001})}\BibitemShut {NoStop}%
\bibitem [{Note1()}]{Note1}%
  \BibitemOpen
  \bibinfo {note} {This algorithm, which performs steepest descent to minimize
  $E$ over $\protect \mathcal {H}$ with cost independent of $d$, should not be
  confused with the steepest descent convex roof algorithm, which minimizes the
  three-tangle over $\Upsilon _\rho $ and which has cost scaling like $d^8
  \protect \qopname \relax o{log}d$.}\BibitemShut {Stop}%
\bibitem [{Note2()}]{Note2}%
  \BibitemOpen
  \bibinfo {note} {These pure states could also be found using a root finding
  algorithm on the polynomial defined in the proof of Theorem~\ref
  {thm1}.}\BibitemShut {Stop}%
\bibitem [{Note3()}]{Note3}%
  \BibitemOpen
  \bibinfo {note} {Caratheodory's theorem implies that pure-state ensembles of
  size $d^2$ are required in general in order to minimize the convex roof. For
  this reason, $\pi _i$ is constructed as a uniform mixture of $3/2 d-i$
  distinct pure states selected from the range of $\rho _i$, each with zero
  $E$. Because $\DOTSB \sum@ \slimits@ _{i=0}^{d-2}(3/2 d-i) = d^2-1$ his
  ensures that $d^2$ pure states are selected in total, including the final
  pure state.}\BibitemShut {Stop}%
\bibitem [{\citenamefont {Eltschka}\ and\ \citenamefont
  {Siewert}(2012)}]{ES12B}%
  \BibitemOpen
  \bibfield  {author} {\bibinfo {author} {\bibfnamefont {C.}~\bibnamefont
  {Eltschka}}\ and\ \bibinfo {author} {\bibfnamefont {J.}~\bibnamefont
  {Siewert}},\ }\href@noop {} {\bibfield  {journal} {\bibinfo  {journal} {Phys.
  Rev. Lett.}\ }\textbf {\bibinfo {volume} {108}},\ \bibinfo {pages} {20502}
  (\bibinfo {year} {2012})}\BibitemShut {NoStop}%
\end{thebibliography}
\end{document}